# Predictability and Prediction for an Experimental Cultural Market

Richard Colbaugh[*], Kristin Glass[†], and Paul Ormerod[‡]

[*]rcolbau@sandia.gov, Sandia National Laboratories, Albuquerque, NM USA
[†]kglass@icasa.nmt.edu, New Mexico Tech, Socorro, NM USA
[‡]pormerod@volterra.co.uk, Volterra Consulting, London, UK

**Abstract.** Individuals are often influenced by the behavior of others, for instance because they wish to obtain the benefits of coordinated actions or infer otherwise inaccessible information. In such situations this social influence decreases the ex ante predictability of the ensuing social dynamics. We claim that, interestingly, these same social forces can *increase* the extent to which the outcome of a social process can be predicted very early in the process. This paper explores this claim through a theoretical and empirical analysis of the experimental music market described and analyzed in [1]. We propose a very simple model for this music market, assess the predictability of market outcomes through formal analysis of the model, and use insights derived through this analysis to develop algorithms for predicting market share winners, and their ultimate market shares, in the very early stages of the market. The utility of these predictive algorithms is illustrated through analysis of the experimental music market data sets [2].

## 1 Introduction

Enormous resources are devoted to the task of predicting the outcomes of social processes, in domains such as economics, public policy, popular culture, and national security, but the quality of such predictions is often quite poor. Consider, for instance, the case of cultural markets. Perhaps the two most striking characteristics of these markets are their simultaneous *inequality*, in that hit songs, books, and movies are many times more popular than average, and *unpredictability*, so that well-informed experts routinely fail to identify these hits beforehand. Examination of other domains in which the events of interest are outcomes of social processes reveals a similar pattern – market crashes, regime collapses, fads and fashions, and "emergent" social movements involve significant segments of society but are rarely anticipated.

It is tempting to conclude that the problem is one of insufficient information. Clearly winners are qualitatively different from losers or they wouldn't be so dominant, the conventional wisdom goes, so in order to make good predictions we should collect more data and identify these crucial differences. Research in the social and behavioral sciences calls into question this conventional wisdom and, indeed, indicates that there may be fundamental limits to what can be predicted about social systems. Consider social processes in which individuals pay attention to what others do. Recent empirical studies offer evidence that the *intrinsic* characteristics of such processes, such as the quality of the various options in a social choice situation, often do not possess much predictive power [3-10].

In order to understand this phenomenon more deeply, Salganik, Dodds, and Watts [1] conducted an elegant experiment in which over 14,000 participants were recruited to participate in an "artificial" music market and the impact of social influence on their choice of songs to download was examined. Briefly, the participants were presented with a web page displaying a selection of 48 songs by unknown bands and were asked to choose songs to listen to and download. As they arrived at the music market site they were randomly assigned to one of two experimental conditions: Independent, in which they saw only the names of bands and songs, and Social Influence, in which they were further divided into eight distinct "worlds" and could see (in addition to the bands and songs) the number of times each song had been downloaded by previous participants in their respective worlds. There were three main findings: 1.) song "quality" is only weakly related to market share success, 2.) the presence of social influence leads to "herding" behavior regarding song popularity, and 3.) increasing the strength of social influence increases both inequality and unpredictability of market outcomes.

Both the empirical analyses [3-10] and the experimental study [1] offer evidence that, for many social processes, it is not possible to obtain useful predictions using standard methods, which focus almost exclusively on the intrinsic characteristics of the process and its possible outcomes. We propose that useful prediction requires consideration of both intrinsics and the underlying *social dynamics*, and present in [11] a new approach to predictive analysis which leverages this idea. This paper applies this analytic framework to the experimental music market described in [1,2] and derives two main results. First, we develop a simple model for cultural markets (such as the music market [1]) which captures both process intrinsics and social influence dynamics. This model is employed to formally assess the predictability of market outcomes for various sets of candidate measurables and thereby identify measurables which possess predictive power. Second, using insights derived from this predictability analysis we formulate algorithms for predicting market share winners, and their ultimate market shares, very early in the market's evolution; the performance of these algorithms is illustrated through predictive analysis of the experimental music market data sets available in [2].

## 2 Predictability Assessment

**Basic idea.** A defining characteristic of cultural markets is that participants are influenced by the behavior of others, for instance because they wish to obtain the benefits of coordinated action (e.g., enjoy the offering with friends) or infer otherwise inaccessible information (e.g., by observing people "in the know"). Processes in which observing a certain behavior increases an individual's probability of adopting that behavior are often referred to as *positive externality processes* (PEP), and we use that term here. One hallmark of PEP is their apparent unpredictability: phenomena from hits in cultural markets to crashes in financial markets to political upheavals appear resistant to predictive analysis (although there is no shortage of *ex post* explanations for their occurrence!).

It is not difficult to gain an intuitive understanding of the basis for this unpredictability. Individual preferences and opinions are mapped to collective outcomes through an intricate, dynamical process in which people react individually to an environment consisting of others who are reacting likewise. Because of this feedback dynamics, collective outcomes can be quite different from those implied by simple aggregations of individual preferences; standard prediction methods, which typically are based (implicitly or explicitly) on and aggregation ideas, do not capture these dynamics and therefore are often unsuccessful. Interestingly, the feedback dynamics which reduces PEP predictability based on simple preference aggregation may *increase* the predictive power of very early measurements of these dynamics. Again the intuition is clear: early trends are reinforced through the positive feedbacks of PEP, suggesting the possibility that early rankings of alternatives may be informative concerning the ultimate outcomes. We now explore this intuition more formally

**Model.** Consider an online market, such as the music market [1], in which individuals visit a web site, browse an assortment of available items, and choose one or more items to download. For simplicity, we focus on a market visited by a sequence of consumers, with each visitor choosing between two items {A, B}; generalizing this simple binary choice setting to any finite number of choices is straightforward [12,13]. We model this situation by supposing that agent i chooses item A with probability

$\Sigma_{online}$ $\qquad\qquad\qquad P_i(A) = \beta\pi + (1-\beta)\,f$

where $f \in [0,1]$ is item A's current market share, $(1-\beta)$ quantifies the intensity of social influence (with $\beta \in [0,1]$), and $\pi$ is the probability of an agent choosing A in the "no social influence" case (i.e., when $\beta=1$). Agent i selects item B with probability $1 - P_i(A)$. In this model, $\pi$ can be interpreted as a measure of the "appeal" of item A (relative to B), f is the social signal, and $\beta$ quantifies the relative importance of appeal and social influence in the decision-making process.

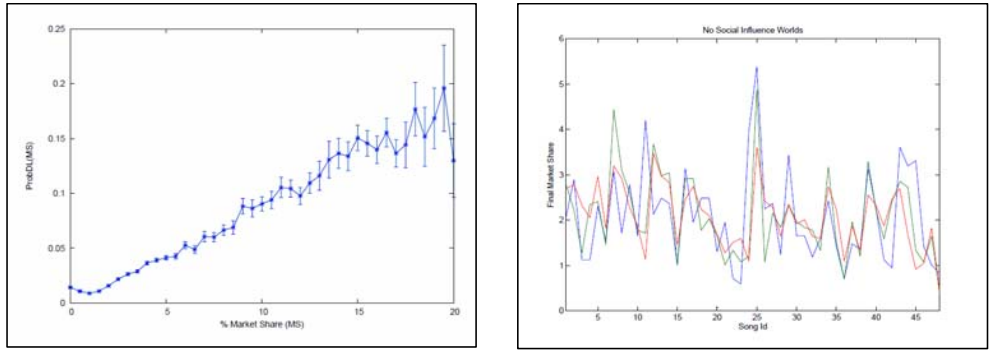

**Fig. 1.** Some characteristics of the music market dynamics. The plot at left shows that in the social influence worlds the market is PEP, with song download probability increasing with number of previous downloads. The plot at right depicts ultimate market share for all songs in the no social influence worlds and indicates that, in the absence of social influence, a few songs are consistently appealing.

The model $\Sigma_{online}$ is extremely simple, perhaps the simplest possible representation which captures the effects of both social influence and appeal in an online market. Nevertheless, this model reflects key behaviors observed in the music market [1] as well as in other cultural markets [e.g., 9]. For instance, Figure 1 provides a quantitative characterization of the roles of social influence and appeal for the music market [1]. The plot at the left of Figure 1 shows that in the Social Influence worlds the market exhibits positive externalities, with song download probability (vertical axis) increasing with number of previous downloads (horizontal axis); the error bars represent two standard errors. The plot at the right of the figure shows ultimate market share for all songs in the Independent worlds and indicates that, in the absence of social influence, some songs (e.g., song 25) are more appealing. Note that, in particular, the dependence of song download probability on previous downloads is approximately linear and capturing intrinsics in terms of download probability in the Independent condition appears reasonable. Moreover, simulations of $\Sigma_{online}$ show that as social influence (SI) increases ($\beta$ decreases) both inequality and unpredictability of market shares increase. Thus, despite its simplicity, $\Sigma_{online}$ provides a useful starting point for studying predictability of online markets. Note that $\Sigma_{online}$ can be written as a set of stochastic differential equations with state variables $x_1 = f$ and $x_2 = 1/(t+1)$ [12], so that the system's predictability properties can be evaluated using the methods presented in [11].

**Predictability assessment.** Consider the predictability of ultimate market share for the system $\Sigma_{online}$. We develop in [11] a mathematically rigorous approach to predictability assessment for a broad range of dynamical systems. Here we apply this assessment methodology in a somewhat informal, intuitive manner; the reader interested in a more formal analysis is referred to [11, 12]. Our main interest in this study is *eventual state (ES) predictability*. Roughly, a system Σ is ES predictable if qualitatively different outcomes (say hit or flop in a cultural market) have sufficiently different probabilities of occurring when Σ is initialized at similar starting configurations (see [11] for a more precise definition). A key aspect of this definition is that it depends the specification for initial configurations, which in turn depends on which system states and parameters are assumed measurable; indeed, in what follows we will use the results of ES predictability assessment to *identify* measurables with predictive power.

The standard approach to market share prediction is to assume that item appeal is a relevant measurable, estimate appeal in some way, and use this estimate to predict market share. To examine the utility of this approach, we assess ES predictability of market share for items with identical appeal ($\pi=1/2$) and identical initial market shares ($f(0)=1/2$). If it is likely that the market will evolve so one or the other item dominates (f becomes large or small), then market dynamics is not very dependant on item appeal and therefore is unpredictable using the standard approach. In this case we should seek a different prediction method, perhaps based on other measurables. Alternatively, if market dominance by either item is unlikely then the market dynamics depends on item appeal in a more predictable way and the standard method may be useful.

We evaluate ES predictability using the assessment procedure proposed in [11]. Let $X_{s1}$ and $X_{s2}$ be two subsets of the state space of $\Sigma_{online}$ corresponding to, respectively, $f \approx 1/2$ (approximately equal market share) and large/small f (market dominance by one or the other item). Define the set of similar initial configurations $X_0$ to be a small set surrounding $f(0) = 1/2$, the identical initial market share condition. Then, if both $X_{s1}$ and $X_{s2}$ are likely to be reached from $X_0$, the problem is ES unpredictable (and also unpredictable in a practical sense). See Figure 1 for a sketch depicting the basic setup.

As an illustration of the insights obtainable with such analysis, consider the high social influence (SI) case corresponding to small β in $\Sigma_{online}$. For a broad range of noise models, the analysis generates fairly high probabilities for reachability of both $X_{s1}$ and $X_{s2}$ from $X_0$. Thus two qualitatively different outcomes – market share equity ($X_{s1}$) and market shares dominance ($X_{s2}$) – are both likely, indicating that the system is ES unpredictable. This result is consistent with empirical findings for cultural markets [e.g., 1] and suggests that the standard approach to market share prediction is not likely to produce accurate forecasts.

Next consider the problem of searching for alternative measurables which provide better predictability properties in the high SI case. For example, it might be supposed that very early market share time series data would be useful for prediction when SI is high. The intuition behind this idea is that the "herding" behavior that can arise from SI, and which makes market prediction hard using standard methods, may lead to a lock-in effect, in which very early market share leaders become difficult to displace. To test this hypothesis, define $X_0^*$ to be a small set surrounding $f(t^*) = 1/2$, where $t^*$ is a small *but nonzero* time (see Figure 2). We compute, using the ES predictability assessment algorithm given in [11], the probability that $\Sigma_{online}$ with $\pi=1/2$ will evolve from $X_0^*$ to $X_{s1}$ and $X_{s2}$. In this case, the analysis yields a high probability of reaching $X_{s1}$ and low probability of reaching of $X_{s2}$ (typical probabilities are on the order ~0.9 and ~$10^{-3}$, respectively). Thus using very early time series data produces a more predictable situation, in which indistinguishable market configurations evolve to indistinguishable outcomes.

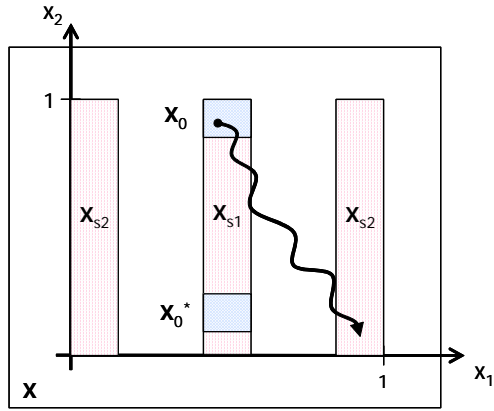

**Fig. 2.** Setup for online market predictability assessment.

## 3 Market Prediction

**Introduction.** The insights obtained through the predictability assessment procedure summarized above can be used to derive methods for predicting outcomes of the evolution of the music market [1]. We are interested in two problems: 1.) identifying market share winners very early in the market's evolution and 2.) predicting ultimate market shares for these winning songs.

**Identifying market share winners.** The objective here is to identify a practically measurable "indicator" condition which enables successful songs to be recognized early in the music market's evolution. We focus on the ten high SI markets in the experiments [2], as these are the most unpredictable using standard methods [1]. Our method is simple and natural. The distribution of downloads in high SI markets is right-skewed, reflecting the PEP nature of these markets and in particular the tendency for market share "lock in" to occur early in the process. This observation suggests that when a market first exhibits signs of right skew in market share distribution, a good prediction for the song that will ultimately win the largest market share is the one with leading market share at that point.

Consider the simple measure of right-skew $MM_i(t) = mean_i(t)/median_i(t)$, where $mean_i(t)$ is market share mean for the 48 songs in (high SI) world i at time t and $median_i(t)$ is the analogous median (time t is measured in market "ticks" [2]). The plot in Figure 3 shows that the dynamics of $MM_i(t)$ provides a reliable early means of distinguishing high SI markets from the low and no SI markets. Moreover, these data indicate that high SI markets reach $MM_i(t) \geq 1.1$ very early (i.e., at ~5% of the total market trajectory). Thus we propose the following method for identifying market share winners: for a given market i, predict as the ultimate market share winner the song with leading market share when $MM_i(t) \geq 1.1$ for the first time. Implementing this strategy with the music market data [2] yields the following results: 1.) the winning song is correctly identified in 90% of the high SI markets and 2.) this identification is made within the first ~5% of the market trajectory.

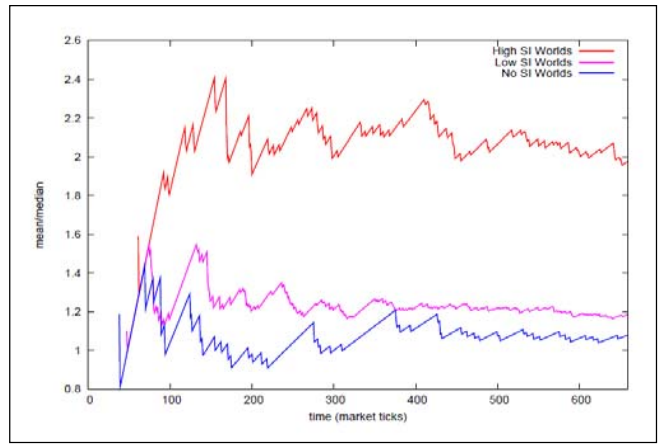

**Fig. 3.** Dynamics of $MM_i(t)$. The plot shows average $MM_i(t)$ for the high SI (red), low SI (magenta), and no SI (blue) music markets.

**Predicting ultimate market shares.** Consider next the problem of predicting the market shares for successful songs early in the market's evolution. More specifically, we wish predict the ultimate market share for the top five songs in a given market. Again the focus is on high SI markets, as these are unpredictable using standard methods [1]. We propose the following very simple prediction model:

$\Sigma_{ms}$:     $ms_i(T) = \kappa + \alpha\, ms_{i,\,noSI} + \beta_1\, ms_i(\tau) + \beta_2\, ms_i(k\tau) + \beta_3\, ms_{i,\,mw}(a\tau, k\tau),$

where $ms_i(t)$ is the market share of song i at time t, T is end time for the market under study, $m_{i,\,noSI}$ is the mean ultimate market share for song i in the no SI markets, $\tau$ is the time at which this market first reaches $MM(t) \geq 1.1$, $ms_{i,\,mw}(t_1, t_2)$ is the mean market share for song i over the "moving window" $[t_1, t_2]$, $k \geq 1$ defines how much early market share time series is available to the prediction model, and $\{a, \kappa, \alpha, \beta_1, \beta_2, \beta_3\}$ are the model parameters. Note that obviously more sophisticated prediction models could be used; here the goal is to demonstrate useful performance with a simple linear regression predictor.

We now summarize some results of applying $\Sigma_{ms}$ to the task of predicting the ultimate market share of 50 successful songs (the top five songs in each of the ten high SI markets). Note first that, for a broad range of early time series availability (i.e., k specified so that 5%–50% of market time series is used for prediction), all terms in $\Sigma_{ms}$ (except $\kappa$) are statistically significant predictors of final market share ($p<0.05$). Next, consider the extent to which final market share can be predicted using only the intrinsic appeal of the songs, as measured by $\alpha\, ms_{i,\,noSI}$. As shown in Figure 4 (plot at left), this quantity has limited predictive power, explaining less than 50% of the variance of final market share over the 50 successful songs. In contrast, the most predictive dynamics term $\beta_2 ms_i(k\tau)$ can provide useful predictions even if only a small amount of early time series is available (e.g., this term explains ~80% of final market share variance if 15% of early time series is available); see Figure 4 (plot at left).

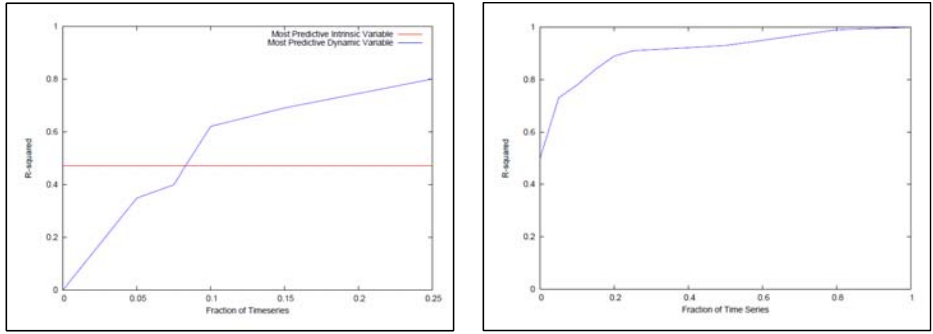

**Fig. 4.** Sample music market prediction results. The plot at left shows the fraction of final market share variance explained by the most predictive intrinsic alone ($\alpha\, ms_{i,\,noSI}$, red) and most predictive dynamics variable alone ($\beta_2\, ms_i(k\tau)$, blue), as a function of the fraction of early time series available. The plot at right depicts the fraction of final market share variance explained by $\Sigma_{ms}$ as a function of the fraction of early time series available.

As expected from the predictability assessment summarized in Section 2, the predictive power of $\Sigma_{ms}$ increases rapidly as a function of amount of early market share time series available to the model; this dependence is shown in Figure 4 (plot at right). Finally, a tenfold cross validation study shows that the model $\Sigma_{ms}$, although simple, provides good out-of-sample prediction performance. For example, with access to the first 15% of market share time series the model provides an out-of-sample prediction accuracy of ~82%.

**Acknowledgements.** This research was supported by the U.S. Department of Homeland Security, the U.S. Department of Defense, and the Laboratory Directed Research and Development program at Sandia National Laboratories.

## References

1. Salganik, M., P. Dodds, and D. Watts, "Experimental study of inequality and unpredictability in an artificial cultural market", *Science*, Vol. 311, pp. 854-856, 2006.
2. www.princeton.edu/~mjs3/data.shtml (accessed 2009).
3. Arthur, W., "Competing technologies, increasing returns, and lock-in by historical events", *Economic Journal*, Vol. 99, pp. 116-131, 1989.
4. Bikhchandani, S., D. Hirshleifer, and I. Welch, "Learning from the behavior of others", *J. Economic Perspectives*, Vol. 12, pp. 151-170, 1998.
5. Hedstrom, P., R. Sandell, and C. Stern, "Mesolevel networks and the diffusion of social movements", *American J. Sociology*, Vol. 106, pp. 145-172, 2000.
6. Shiller, R., *Irrational Exuberance*, Princeton University Press, NJ, 2000.
7. Rogers, E., *Diffusion of innovations*, Fifth Edition, Free Press, NY, 2003.
8. Walls, W., "Modeling movie success when 'nobody knows anything': Conditional stable-distribution analysis of film returns", *J. Cultural Economics*, Vol. 29, pp. 177-190, 2005.
9. Colbaugh, R. and K. Glass, "Predictability and prediction of social processes", *Proc. 4th Lake Arrowhead Conference Human Complex Systems*, Lake Arrowhead, CA, April 2007.
10. Easley, D. and J. Kleinberg, *Networks, Crowds, and Markets: Reasoning About a Highly Connected World*, Cambridge University Press, 2010 (to appear).
11. Colbaugh, R. and K. Glass, "Predictive analysis for social processes I: Multi-scale hybrid system modeling, and II: Predictability and warning analysis", *Proc. 2009 IEEE Multi-Conference on Systems and Control*, Saint Petersburg, Russia, July 2009.
12. Colbaugh, R. and K. Glass, "Predictive analysis for social processes", Sandia National Laboratories SAND Report 2009-0584, January 2009.
13. Ormerod, P. and R. Colbaugh, "Cascades of failure and extinction in evolving complex systems", *J. Artificial Societies Social Simulation*, Vol. 9, No. 4, 2006.